\documentclass[prd, superscriptaddress ,amsmath,amssymb]{revtex4}
\usepackage{latexsym}
\usepackage{amssymb}
\usepackage{epsfig,amsmath,graphics}
\usepackage{epstopdf}
\usepackage{verbatim}

\newcommand{\OO}{\mathcal{O}}

\newcommand{\MeV}{\text{MeV}}
\newcommand{\cm}{\text{cm}}
\newcommand{\Hz}{\text{Hz}}

\newcommand{\Bext}{B_\text{ext}}
\newcommand{\vBext}{\vec{B}_\text{ext}}
\newcommand{\tshot}{t_\text{shot}}

\begin{document}

\title{Cosmic Axion Spin Precession Experiment (CASPEr)}

\author{Dmitry Budker}
\affiliation{Department of Physics, University of California, Berkeley, California 94720, USA}
\affiliation{Nuclear Science Division, Lawrence Berkeley National Laboratory, Berkeley, California 94720, USA}

\author{Peter W. Graham}
\affiliation{Stanford Institute for Theoretical Physics, Department of Physics, Stanford University, Stanford, CA 94305}

\author{Micah Ledbetter}
\affiliation{AOSense, 767 N. Mary Ave, Sunnyvale, CA, 94085-2909}

\author{Surjeet Rajendran}
\affiliation{Stanford Institute for Theoretical Physics, Department of Physics, Stanford University, Stanford, CA 94305}

\author{Alexander O. Sushkov}
\affiliation{Department of Physics, and Department of Chemistry and Chemical Biology, Harvard University, Cambridge, MA 02138, USA.}

\begin{abstract}
We propose an experiment to search for QCD axion and axion-like-particle (ALP) dark matter.
Nuclei that are interacting with the background axion dark matter acquire time-varying CP-odd nuclear moments such as an electric dipole moment.
In analogy with nuclear magnetic resonance, these moments cause precession of nuclear spins in a material sample in the presence of an electric field.
Precision magnetometry can be used to search for such precession.
An initial phase of this experiment could cover many orders of magnitude in ALP parameter space beyond the current astrophysical and laboratory limits.
And with established techniques, the proposed experimental scheme has sensitivity to QCD axion masses $m_a \lesssim 10^{-9}$ eV, corresponding to theoretically well-motivated axion decay constants $f_a \gtrsim 10^{16}$ GeV.
With further improvements, this experiment could ultimately cover the entire range of masses $m_a \lesssim \mu$eV, complementary to cavity searches.
\end{abstract}

\maketitle

\section{Introduction}
The discovery of the nature of dark matter would provide significant insights into particle physics, astrophysics, and cosmology.
While the Weakly Interacting Massive Particle (WIMP) is a well motivated candidate, it is heavily constrained by null results from a variety of experiments \cite{Ahmed:2011gh, Aprile:2011ed, Akerib:2013tjd}.
Further, the Large Hadron Collider has placed stringent constraints on scenarios such as supersymmetry that have provided the theoretical basis for WIMP dark matter \cite{Lowette:2012uh}.
Indeed, these constraints are most easily alleviated by allowing for a rapid decay of the supersymmetric WIMP candidate (e.g.~\cite{Graham:2012th}),  precluding a cosmological role for it.
Thus, it is essential to develop techniques to search for a wide class of dark matter candidates.

Introduced as a solution to the strong CP problem \cite{Peccei:1977hh, Peccei:1977ur}, the axion is a prominent dark matter candidate. It arises naturally as the pseudo Goldstone boson of some global symmetry that
is broken at a high scale $f_a$ \cite{Weinberg:1977ma, Wilczek:1977pj, Kim:1979if, Shifman:1979if, Dine:1981rt, Zhitnitsky:1980he}. QCD generates a potential  $\frac{1}{2} m_a^2 a^2$ for the axion with  $m_a \sim \frac{\Lambda^{2}_{\text{QCD}}}{f_a}$. An initial displacement of the axion field from its minimum results in oscillations of this field  with frequency $m_a \frac{c^2}{\hbar}$. The energy density in these oscillations can be dark matter  \cite{Preskill:1982cy, Dine:1982ah}.
Other types of light bosons, often called axion-like-particles (ALPs), have attracted significant attention~\cite{Hewett:2012ns, Ringwald:2012cu, Ringwald:2012hr, Baker:2011na, Arias:2010bh, Jaeckel:2010ni, Ehret:2010mh, Schott:2011fm, Battesti:2010dm, Conlon:2006tq, Arvanitaki:2009fg, Acharya:2010zx, Pospelov:2012mt, ALP space}. These receive a potential (and a mass) from non-QCD sources and are less constrained than the QCD axion. Like the oscillations of the QCD axion, oscillations of the ALP field in its potential can also be dark matter. We focus on light ALPs with masses $m_a$ comparable to that of the axion.  We will use the term ALP to refer to any of these light bosons, including the QCD axion. The temporal coherence of the oscillations of the dark matter ALP field in an experiment is limited by motion through the spatial gradients of the field. The size of these gradients is set by the de-Broglie wavelength, giving rise to a coherence time $\tau_a \sim \frac{2 \pi}{m_a v^2} \sim 10^6 \frac{2 \pi}{m_a}$, where $v \sim 10^{-3}$ is the galactic virial velocity of the ALP dark matter \cite{ALP space}.

The axion's properties are determined by  $f_a$. Astrophysical bounds rule out axions with $f_a \lesssim 10^{10}$ GeV \cite{Raffelt:2006cw}.
While $f_a \gtrsim 10^{12}$ GeV used to be claimed to be ruled out by cosmological arguments, this was based on a simplified picture of cosmology and is not a rigorous bound (see for example \cite{Linde:1987bx, ALP space}).
The conversion of axions into  photons in the presence of a magnetic field can be used to search for axions with $f_a \sim 10^{12}$ GeV   \cite{Sikivie:1985yu, Asztalos:2009yp}, but the ability of such techniques to probe axions with $f_a \gg 10^{12}$ GeV is limited. It is  important to develop techniques that can search for axions over the vast majority of parameter space up to $f_a \sim  10^{19}$ GeV, especially because of the generic theoretical expectation that the symmetry breaking scale $f_a$ should be close to other fundamental scales in particle physics such as the grand unified  ($\sim 10^{16}$ GeV) and Planck ($\sim 10^{19}$ GeV) scales \cite{Svrcek:2006yi}.

The axion field induces a time varying nucleon electric dipole moment (EDM) $d_n  \sim  10^{-16} \, \frac{a_0 \, \cos \left(m_a t\right)}{f_a} \, \text{e} \cdot \text{cm}$ \cite{Graham:2011qk, ALP space}.  Here $a_0$ is the local amplitude of the axion dark matter field. See \cite{ALP space} for detailed formulas and derivations of these results.  This EDM is generated from the defining coupling $ \frac{a}{f_a} \text{tr } G \tilde{G}$ of the axion to QCD \cite{Weinberg:1977ma, Wilczek:1977pj, Kim:1979if, Shifman:1979if, Dine:1981rt, Zhitnitsky:1980he},  caused by the same QCD dynamics that leads to physical effects for the operator  $ \theta_{\text{QCD}} \, \text{tr } G \tilde{G}$ ({\it e.g.} nucleon EDMs $d_n  \sim  10^{-16} \, \theta_{\text{QCD}} \, \text{e} \cdot \text{cm}$) \cite{Pospelov:1999ha}, resulting in the strong CP problem and its resolution, the axion.  Essentially, the dark matter axion can be thought of as an oscillating value of $\theta_{\text{QCD}}$.

 All EDM experiments to date have searched for static EDMs and have greatly reduced sensitivity to the oscillating nuclear EDM induced by an axion. But in fact, the oscillation of the EDM should, in many ways, make searches easier. Even though the axion is generated by physics at high energies ($f_a \gg 10^{11}$ GeV), its ultra-light mass lies at frequencies accessible in the laboratory. A signal that naturally oscillates at a frequency set by fundamental physics, independent of the details of any particular experiment, should ameliorate many of the systematic errors that often limit the sensitivity of EDM searches.

It was pointed out in \cite{Graham:2011qk} that axion dark matter could be detected in future molecular interferometers using this oscillating EDM.  Here we argue that such an oscillating EDM can be observed through solid-state NMR-based experiments using presently available technology.  We further exploit the oscillatory nature of the signal by designing a resonant detector that enhances the signal, potentially allowing detection of the QCD axion. The nucleon EDM naturally induced by an axion is the primary focus of this paper. However, such interactions may also exist for ALPs and our techniques will also search for them \cite{ALP space}.

\section{Experimental Concept}
``Solid-state EDM" experiments \cite{Shapiro, Mukhamedjanov:2004ki, Budker2, Sushkov1, Rushchanskii1} have been proposed as promising ways to search for static EDMs of electrons and nucleons, and an experimental limit on the electron EDM has been set using these methods \cite{Eckel1}.  This result was not competitive with the current best limit on the electron EDM, due to the systematic effect of sample heating caused by electric field reversal in a dissipative ferroelectric material. We propose an experiment that uses the solid-state approach, together with magnetic-resonance techniques, to search for axion or ALP dark matter. Crucially, since the nucleon EDM is intrinsically time varying, unlike in static EDM searches, it can be detected without electric field reversals. This eliminates systematics that plagued the solid-state EDM experiments.

\begin{figure}
\begin{center}
\includegraphics[width=3.5 in]{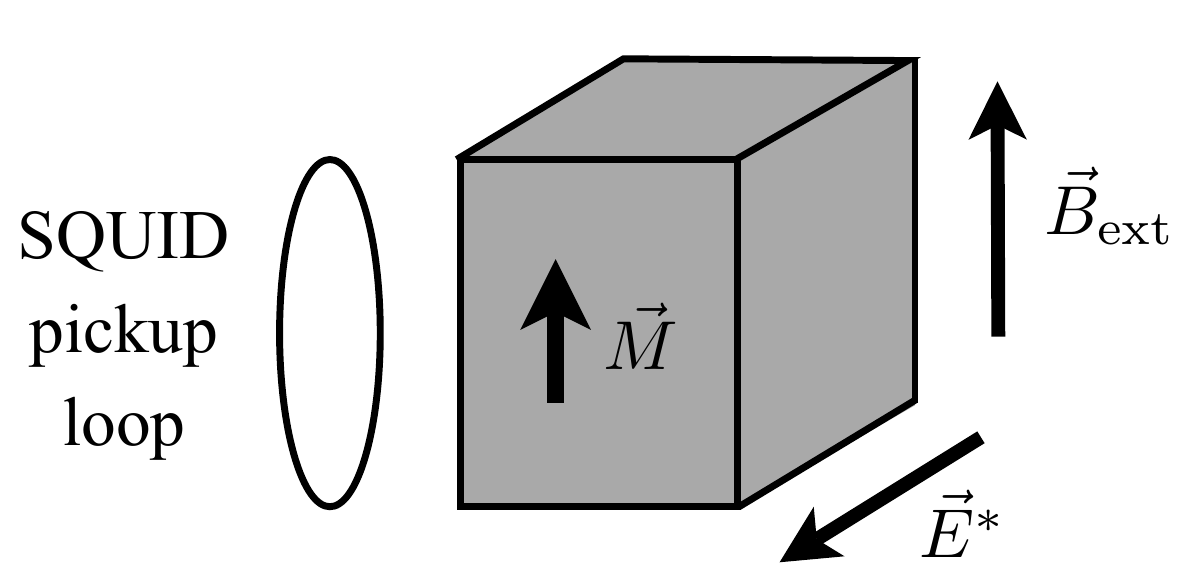}
\caption{ \label{Fig:setup} Geometry of the experiment.  The applied magnetic field $\vec{B}_\text{ext}$ is colinear with the sample magnetization, $\vec{M}$.  The effective electric field in the crystal $\vec{E}^*$ is perpendicular to $\vBext$.  The SQUID pickup loop is arranged to measure the transverse magnetization of the sample.}
\end{center}
\end{figure}

Nuclear spins in a solid insulating material are pre-polarized and placed in an external magnetic field $\vBext$, with an electric field ($\vec{E}^*$) applied perpendicular to $\vBext$, as in Fig.~\ref{Fig:setup}. In the rotating frame, in which $\vBext$ is eliminated, if there is a nucleon EDM, the nuclear spins precess around the electric field.
This results (as seen in the lab frame) in a magnetization at an angle to $\vBext$, which precesses around this field with Larmor frequency.
This transverse magnetization can be measured with a magnetometer such as a superconducting quantum interference device (SQUID) with a pickup loop oriented as shown in Fig.~\ref{Fig:setup}.
For a static EDM the transverse magnetization will not build up in time since its direction relative to the electric field continually oscillates.
Likewise, when the ALP-induced EDM oscillation frequency is different from the Larmor frequency, no measurable transverse magnetization ensues. However, when the two frequencies coincide, there occurs a resonance akin to that in the usual NMR.
The magnitude of the external magnetic field ($\Bext$) is swept to search for this resonance.  At time $t=0$ the spins are prepared along $\vBext$, at subsequent times the magnitude of the transverse magnetization is given by
\begin{equation}
\label{eqn: magnetization signal}
M(t) \approx n p \mu E^* \epsilon_S d_n \frac{\sin \left[ \left( \frac{2 \mu \Bext - m_a c^2}{\hbar} \right) t \right]}{\frac{2 \mu \Bext - m_a c^2}{\hbar}}  \sin \left(  2 \mu \Bext t \right ),
\end{equation}
where $n$ is the number density of nuclear spins, $p$ is the polarization, $\mu$ is the nuclear magnetic dipole moment, and we assume a spin-1/2 nucleus.  Technically, by Schiff's theorem, there can be no net electric field at the nucleus, so the effect of the EDM is actually zero.  Instead the signal actually arises from the Schiff moment.  Following standard convention, we parametrize this effect as $\epsilon_S$, the Schiff suppression factor \cite{Khriplovich:1997ga}, times $d_n$, the magnitude of the ALP-induced nuclear EDM.  Thus $\epsilon_S d$ acts as the effective EDM of the nucleus in the material and $d_n \epsilon_s E^*$ is the energy shift produced between spin-up and spin-down states of the nucleus.  The resonant enhancement occurs when $2 \mu \Bext \approx m_a c^2$.

The nuclear magnetic moments in the sample are polarized using a large applied magnetic field ($B_0\approx 10$~T) at low temperature ($\theta_0\approx4$~K), achieving a polarization fraction $\sim 10^{-3}$.  Higher polarizations may be achievable using optical-pumping techniques. The polarization persists for time, $T_1$, set by the spin-lattice relaxation.  At cryogenic temperature, $T_1$ can reach many hours or longer~\cite{Bouchard1}.  It is advantageous to use an element whose nuclear spin is 1/2, which usually leads to longer spin-lattice relaxation times.

On resonance, the net transverse magnetization precesses at the Larmor frequency ($2 \mu \Bext$), as in Eq.~\eqref{eqn: magnetization signal}.
The amplitude of the resonant transverse magnetization increases linearly with time, in principle up to the ALP coherence time $\tau_a$.
In practice this increase may be cutoff earlier if other effects broaden the resonance.  For example, the transverse relaxation time of the nuclear spins, $T_2$, may be shorter than $\tau_a$.
%
%
In this case, $T_2$ would set the maximum resonant enhancement achievable, so that the factor $\frac{\hbar \sin \left[ \left( 2 \mu \Bext - m_a c^2 \right) \hbar^{-1}  t \right]}{2 \mu \Bext - m_a c^2}$ from Eq.~\eqref{eqn: magnetization signal} would have a maximum $\approx T_2$ (in polycrystalline samples at high magnetic fields, the chemical shift anisotropy may broaden the resonance even further).
The magnetic dipole-dipole interaction between nuclear spins sets $T_2 \sim 1$ ms, however dynamic-decoupling schemes have been shown to suppress broadening due to chemical shifts and increase $T_2$ substantially \cite{Dong}.
$T_2$ in excess of 10 s or even 1000 s has been achieved in other materials, for example \cite{Dong, Ledbetter:2012ch, ledbetter2001}.

A material with a crystal structure with broken inversion symmetry at the site of the high-{\it Z} atoms is necessary for generation of a large effective electric field $E^*$, which is proportional to the displacement of the heavy atom from the centro-symmetric position in the unit cell~\cite{Kuenzi:2002bu}. In a ferroelectric, this displacement can be switched by an applied voltage, however, given the oscillating nature of the ALP-induced signal, it may not be necessary to modulate this displacement, in which case any polar crystal can be used. For ferroelectric PbTiO$_3$, the effective electric field is $E^* \approx 3\times 10^8$~V/cm \cite{Mukhamedjanov:2004ki}.  For other materials, where polarization is permanent, this may be higher by a factor of a few. A detailed discussion of the requirements for the sample material is in the Supplemental Materials.

The measurement procedure is as follows.  The sample is repolarized after every time interval $T_1$.  Then the applied magnetic field is set to a fixed value, which must be controlled to a precision equal to the fractional width of the resonance.  The magnetic field value determines the ALP frequency to which the experiment is sensitive.  The transverse magnetization is measured as a function of time with fixed applied magnetic field.   We call a measurement at a given value of magnetic field ``a shot.''
The total integration time at any one magnetic field value, $\tshot$, is set by the requirement that an $\mathcal{O}(1)$ range of frequencies is scanned in 3 years.  If $T_2$ is longer than the ALP coherence time $\tau_a$, then when searching at frequency $m_a \frac{c^2}{\hbar}$ the width of the frequency band is $\approx 10^{-6} \, m_a \frac{c^2}{\hbar}$.  If $T_2$ is shorter than $\tau_a$ then the width of the frequency band is $\sim \frac{\pi}{T_2}$.  Thus we take $\tshot =\frac{10^8 \text{s}}{\text{min}(10^6, \frac{m_a c^2 T_2}{\pi \hbar} ) }$.  Using the magnetization measurements taken over $\tshot$ the power in the relevant frequency band around $\frac{2 \mu \Bext}{\hbar}$ is found.
The applied magnetic field is then changed to the next frequency bin and the procedure is repeated.
The signal of an ALP would be excess power in a range of magnetic fields (ALP frequencies).  If multiple ALPs existed they would appear as multiple spikes at different frequencies.

Note that at the lowest frequencies $\lesssim T_2^{-1}$ the resonance is broadened significantly so that an $\OO(1)$ range of frequencies is covered in any given frequency bin.  In this regime one may use any of the established techniques searching for static nuclear EDMs but with short sampling times $\lesssim \frac{\hbar}{m_a c^2}$, then look for an oscillating signal in the data.

This search for a time varying EDM is substantially different from searches for a static EDM using solid state systems. In searching for a static EDM, it is necessary to separate the energy shift induced by the EDM from other systematic effects. This is accomplished by searching for energy shifts that modulate linearly with the applied electric field in the sample. However, the modulation of the electric field can induce additional systematic shifts in the system that occur at that modulation frequency, competing with the static EDM signal \cite{Eckel1}. This is not the case for a time varying EDM. The ALP induced EDM oscillates at a frequency set by fundamental physics and leads to observable effects in a system whose parameters are static. The time variation provides the handle necessary to separate this signal from other systematic energy shifts and the signal can be detected without the need for additional handles such as electric field reversals. This  eliminates the systematic problems encountered by solid state static EDM searches such as the dissipation effects in the solid material associated with electric field reversals \cite{Eckel1}. 

\section{Sensitivity}
The experimental sensitivity is likely to be limited by the magnetometer, rather than by the backgrounds discussed below.  We assume a SQUID magnetometer with sensitivity $10^{-16} \frac{\text{T}}{\sqrt{\Hz}}$ as calculated from \cite{Lamoreaux:2001hb} for a $\sim 10$ cm diameter sample and pickup loop (see Supplemental Materials).
The sensitivity could be improved with better SQUIDs, a larger sample/pickup loop (see Supplemental Materials), or other types of magnetometers.
For example, atomic SERF magnetometers could potentially improve this by another order of magnitude \cite{Dang:2009jv, Romalis magnetometer}.


\begin{figure}
\begin{center}
\includegraphics[width=\columnwidth]{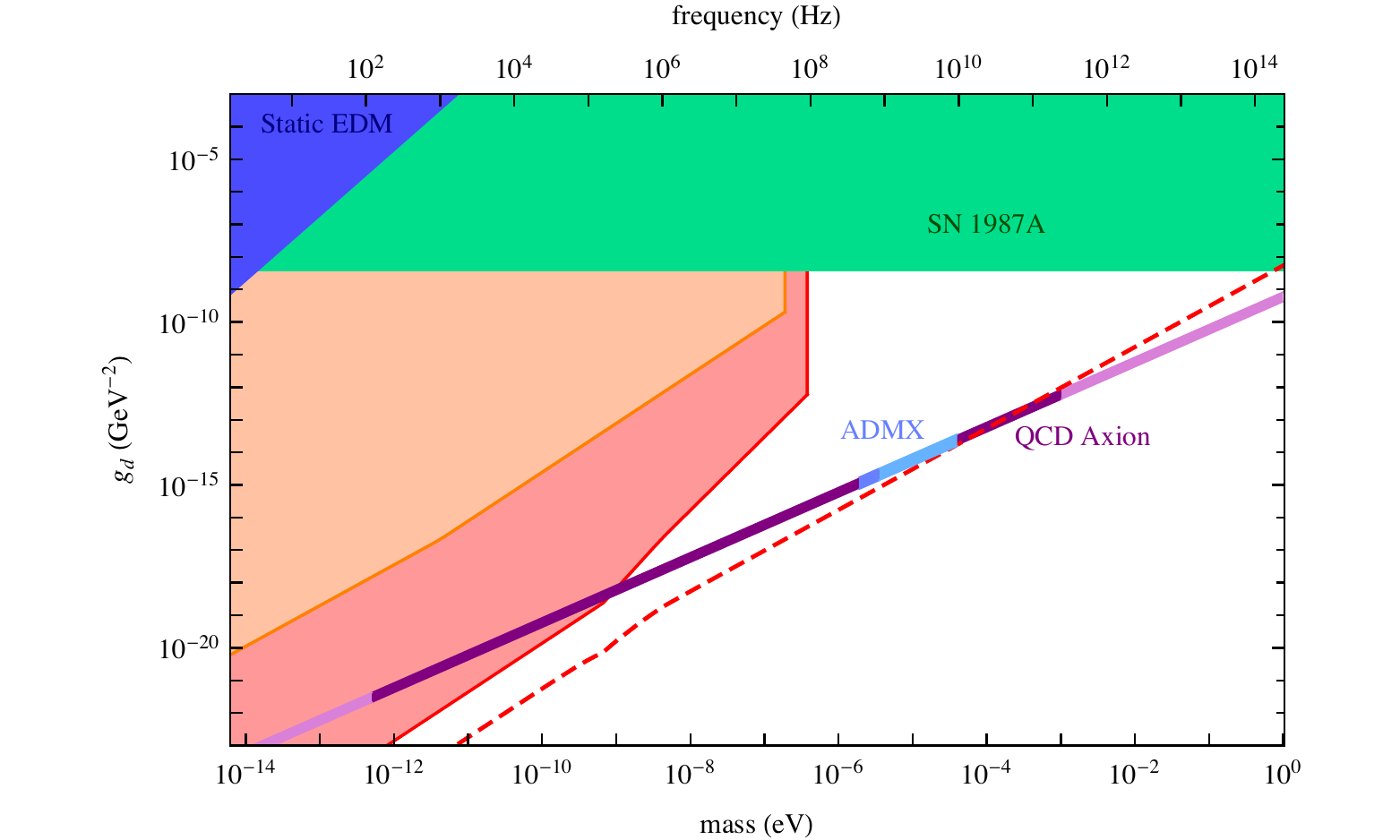}
\caption{ \label{Fig:EDM}  Estimated constraints in the ALP parameter space in the EDM coupling $g_d$ (where the nucleon EDM is $d_n = g_d a$ and $a$ is the local value of the ALP field) vs.~the ALP mass \cite{ALP space}.  The green region is excluded by the constraints on excess cooling of supernova 1987A \cite{ALP space}.  The blue region is excluded by existing, static nuclear EDM searches \cite{ALP space}.  The QCD axion is in the purple region, whose width shows the theoretical uncertainty \cite{ALP space}.  The solid red and orange regions show sensitivity estimates for our phase 1 and 2 proposals, set by magnetometer noise.  The red dashed line shows the limit from magnetization noise of the sample for phase 2.  The ADMX region shows what region of the QCD axion has been covered (darker blue) \cite{Asztalos:2009yp} or will be covered (lighter blue) \cite{ADMXwebpage, snowdarktalk}. Phase 1 is a modification of current solid state  static EDM techniques that is optimized to search for a time varying signal and can immediately begin probing the allowed region of ALP dark matter. To calculate limits from previous (static) EDM searches as well as our sensitivity curves, we assume the ALP is all of the dark matter.
}
\end{center}
\end{figure}

Figure \ref{Fig:EDM} shows the ALP parameter space of the EDM coupling $g_d$ versus ALP mass.  This coupling is defined such that the oscillating nucleon EDM is $d_n = g_d a$ where $a$ is the local value of the classical ALP field (see \cite{ALP space} for a detailed formula).  This is different from the usual ALP-photon coupling parameter.
The purple region of Fig.~\ref{Fig:EDM} shows where the QCD axion lies in this parameter space.  The dark purple is where the QCD axion may be the dark matter.  This parameter space is described in detail in \cite{ALP space}.

The solid (orange and red) regions in Fig.~\ref{Fig:EDM} show estimates for the sensitivities for two phases of our proposed experiments.  Phase 1 (upper, orange region) is a more conservative version relying on demonstrated technology.  Phase 2 (lower, red region) relies on technological improvements which have been demonstrated individually but have not been combined in a single experiment.  Thus the phase 2 proposal may be taken as an estimate of one way to achieve the sensitivity necessary to see the QCD axion with this technique.  Since this is a resonant experiment and the frequency must be scanned, realistically it would likely take several experiments to cover either region.

The dashed (red) line in Fig.~\ref{Fig:EDM} shows the ultimate limit on the sensitivity of the phase 2 experiment from sample magnetization noise (calculated in the Supplemental Materials), which could be reached if the magnetometer is improved.  The sample magnetization noise limit for the phase 1 experiment is not shown, but was calculated and is not a limiting factor for phase 1.  Note the phase 2 noise is small enough that it would not hinder detection of the QCD axion over the entire relevant frequency range.

\begin{table}
\renewcommand{\arraystretch}{1.1}
\begin{center}
\begin{tabular}{|l|c|c|c|c|c|c|}
\hline
& $n$ & $E^*$ & $p$  & $T_2$ & Max. $\Bext$  \\
\hline
Phase 1 & $10^{22} \, \frac{1}{\cm^3}$ & $3 \times 10^8 \frac{\text{V}}{\cm}$ & $10^{-3}$  & 1 \text{ms} & 10 T \\
Phase 2 & $10^{22} \, \frac{1}{\cm^3}$  & $3 \times 10^8 \frac{\text{V}}{\cm}$ & 1 & 1 \text{s} & 20 T \\
\hline
\end{tabular}
\caption{\label{Tab: phases} Parameters for Phase 1 and 2 regions in Figure \ref{Fig:EDM}.}
\end{center}
\end{table}

For both phases we assumed the nucleus is $^{207}\text{Pb}$ so that $\epsilon_s \approx 10^{-2}$ and the nuclear magnetic moment is $\mu = 0.6 \mu_N$, where $\mu_N = 3.15 \times 10^{-14} \, \frac{\MeV}{\text{T}}$.
Other parameters are shown in Table \ref{Tab: phases}.
With these parameters, the limit on the sensitivity of both phase 1 and phase 2 experiments is set by the magnetometer sensitivity.
The upper limit on the ALP mass for the solid curves in Fig.~\ref{Fig:EDM} comes from requiring that the Larmor frequency be less than the maximum achievable frequency using a 10 T (phase 1) or 20 T (phase 2) applied B-field.  The change in slope in the solid phase 2 sensitivity curve comes when $\tau_a = T_2$.

Phase 1 can cover a large piece of unexplored ALP parameter space.  Phase 2 reaches the QCD axion for coupling constants $f_a \gtrsim 10^{16}$ GeV.  If the magnetometer is improved and the magnetization noise limit reached, the QCD axion could possibly be detected over the entire region $f_a \gtrsim 3 \times 10^{13}$ GeV. Significant technological challenges have to be overcome before this experiment reaches the ultimate sensitivity goals of Phase 2. The technological challenges of Phase 1 are relatively easier to overcome and it can immediately begin cutting into ALP dark matter in the allowed region of parameter space in  Fig. \ref{Fig:EDM} \cite{ALP space}. In fact, modifications of current EDM techniques that are optimized to search for a time varying signal can also begin constraining this parameter space. 



\section{Noise Sources}
Transverse, time-varying magnetic fields that vary at frequencies in the measurement bandwidth are a source of noise. The fundamental source of noise is due to the quantum projection of each spin along the transverse directions causing a transverse magnetization of the sample \cite{Budker2,SpinNoise}. 
This noise can be decreased by using larger sample volumes.
 Note that the sensitivity limit set by this fundamental magnetization noise (red dashed line) in Fig. \ref{Fig:EDM} would allow detection of the QCD axion over the entire frequency range accessible in this experiment.

A more detailed discussion of this projection noise, as well as other, technical, sources of experimental noise, such as vibrations, is included in the Appendix. Choosing an optimal experimental strategy  requires engineering studies that are left for future work. The static EDM searches require control over the dc components of these magnetic fields, while our experiment requires control over the ac component at relatively high frequencies $\gtrsim$ kHz. Such control should be easier to achieve and it thus seems plausible that these sources of noise can be adequately suppressed. For example, control over ac magnetic-field noise at comparable levels in the background of a large external field has been demonstrated in current cavity searches for the axion \cite{Asztalos:2009yp}.

\section{Conclusions}
The proposed experiment appears to have sufficient sensitivity to detect axion dark matter, especially when $f_a$ is in the theoretically favored range between the grand unified and Planck scales. It is also sensitive to ALPs that couple to the nucleon EDM, probing sources of symmetry breaking in the ALP sector. The signal in this solid-state NMR-based experiment benefits from the large number ($\gtrsim 10^{22}$) of spins in a solid state system. In conjunction with precision magnetometers, this approach enables sensitivity to this region of parameter space using current technology. In fact, with some improvements in magnetometer sensitivity, this technique could be used to detect the QCD axion with $f_a$ essentially all the way down to the region that can be probed by microwave cavity experiments such as ADMX.  Together then, these techniques may cover almost the entire QCD axion dark matter range.

We do not know of any other approach that can be sensitive to this region of parameter space using present technology. Previous proposals that were aimed at searching for the time varying EDM induced by ALP dark matter were focussed on detecting them through interferometry in ultra-cold molecular systems. These proposals require significant technology development to reach the sensitivity necessary to detect QCD axion dark matter, primarily because they are limited by the substantially smaller effective number of such cold molecules ($\sim 10^6$) that can be produced with current technology. 

In contrast to searches for the dark-matter axion through its coupling to photons, the sensitivity in the present approach has only a weak dependence on $f_a$. This is because the measured spin precession probes the matrix element ($\propto f_a^{-1}$) of the axion interaction. The EDM arises from a non-derivative interaction of the axion and its physical effects are not suppressed by the ratio of the size of the experiment and the axion wavelength. Indeed, the axion dark matter  induced EDM has a fixed magnitude independent of $f_a$.  Since the axion coherence time grows with $f_a$, $\tau_a \propto f_a$, the sensitivity of this experiment also grows with $f_a$.  By contrast experiments relying on the axion-photon coupling $\frac{a}{f_a} F \tilde{F}$ are suppressed by the large axion wavelength at high $f_a$, because $F \tilde{F}$ is a total derivative so all physical effects must be proportional to derivatives of the axion field.

The time varying nature of the axion (or ALP) induced EDM is a key feature of this concept. The EDM oscillates at a frequency set by particle physics, independent of the experimental setup.  This distinguishes the signal from many possible backgrounds and should ameliorate the challenges faced by static EDM searches. For example, electric field reversals are not needed to see this signal, eliminating the  dissipative sample-heating systematics that limited the solid-state static EDM searches. Further, control over noise sources is only required over the signal's relatively high frequency range (kHz - MHz) and narrow bandwidth ($\sim 10^{-6} \, m_a \frac{c^2}{\hbar}$).  Finally, though the induced EDM is small, its oscillation at laboratory frequencies enables resonant schemes that improve the signal to noise ratio. 



%


A positive signal can be verified in many ways. The signal should change with the relative orientation of the nuclear spin and the electric field;  it should  be in phase with that from another sample that is placed within the ALP de Broglie wavelength ($\gg$ 300 km). If there is evidence of a signal at any particular frequency, the apparatus can be tuned to that frequency to determine if the signal builds up in that band as expected.


Similarly to the state of WIMP direct detection at its beginning, there are serious technical challenges that must be overcome for our proposed axion direct-detection technique to reach its ultimate sensitivity.  For example, while there are materials (such as PbTiO$_3$, and solid solutions Pb(Zr,Ti)O$_3$, (Pb,La)(Zr,Ti)O$_3$) that have the properties necessary to begin implementing this proposal, further work is necessary to find an optimal choice.  But, the case for axion (and ALP) dark matter is strong enough to merit the necessary effort. In fact, well before reaching the ultimate sensitivity necessary to see QCD axion dark matter, current EDM techniques can be optimized to search for a time varying EDM that can already search for ALP dark matter in the allowed region of parameter space in  Fig. \ref{Fig:EDM} \cite{ALP space}. With our resonant scheme, the ultimate sensitivity of this proposal allows for the detection of  QCD axion dark matter. As with WIMP detectors, this is a scalable experiment with several possible avenues for improvements in sensitivity.  Reaching the QCD axion requires overcoming significant technological challenges.  But long before that level of sensitivity is reached, this experiment will probe large regions of ALP parameter space beyond current astrophysical and laboratory limits.

A discovery in such an experiment would have profound consequences for physics since it would reveal not just the nature of dark matter and establish the axion as the solution to the strong CP problem, but also provide a window into some of the highest energy scales in nature.

\section*{Acknowledgements}
We would like to thank Blas Cabrera, Savas Dimopoulos, Matt Pyle, and Scott Thomas for valuable discussions.  SR was supported by ERC grant BSMOXFORD no. 228169. This work has been supported in part by the National Science Foundation under grant  PHY-1068875. D.B. acknowledges support by the Miller Institute for Basic Research in Science.  PWG acknowledges the support of NSF grant PHY-1316706, the Hellman Faculty Scholars program, and the Terman Fellowship.

\appendix

\section{Supplemental Materials}

\subsection{Choice of sample material}

The choice of sample material is complicated as there are several tradeoffs.  We do not propose a particular material here.  Instead we discuss the factors which are involved in this decision and illustrate with examples.  The material should be chosen to optimize $\epsilon_s$, $p$, the transverse relaxation time $T_2$, and $E^*$.  Similarly to the situation in WIMP direct detection, many different detector materials may be worth exploring on the path toward detection of axion dark matter.

The sample should be an insulating material containing no unpaired electron spins.  The signal is due to the ALP-induced nuclear Schiff moment which scales with $\epsilon_s \propto Z^3$ ($Z$ is atomic number) \cite{OPSushkov1}.  So the material should have large-Z  atoms whose nuclei have a non-zero magnetic dipole moment, such as $^{207}$Pb that has nuclear spin 1/2.  Heavy elements, e.g.~Pb or Hg have Schiff suppression factors $\epsilon_s \sim 10^{-2}$.  Elements exist with even higher Schiff moments $\epsilon_s \sim 1$, for example the light actinides \cite{Auerbach:1996zd, Spevak:1996tu}, but these are challenging to work with because they are radioactive.

\subsection{Experimental noise sources}

An important source of time varying magnetic fields is $\vec{B}_{\text{ext}}$, the magnetic field that is applied to the sample to tune its Larmor frequency close to the ALP's mass. Transverse fluctuations of this field are a source of noise. There are several strategies that can address them. First, field fluctuations caused by the relative mechanical motion of the sample in relation to $\vec{B}_{\text{ext}}$ can be minimized  by rigidly attaching the sample to the source of the magnetic field. It may also be possible to engineer vibration-isolation systems to damp such sources of noise at the frequencies of interest  to this experiment ($\gtrapprox$ kHz). Second, superconducting magnets can be used to provide $\vec{B}_{\text{ext}}$, significantly suppressing ac field noise from $\vec{B}_{\text{ext}}$. Third, the phase of the signal depends upon the relative orientation between the nuclear spin (the direction of the EDM) and the internal electric field in the crystal. A differential measurement between two samples where these orientations are different will retain the ALP signal while canceling common-mode magnetic-field noise. With this strategy, it is only the gradients of $\vec{B}_{\text{ext}}$ that contribute to the differential signal. These gradients can be minimized through the use of well known coil arrangements such as Maxwell coils. It should be noted that in such coil arrangements, the transverse component of the applied field is also similarly suppressed. Consequently, with this differential-measurement strategy, this noise is suppressed by three potentially small parameters: the initial size of the time variation, the longitudinal gradient, and the size of the transverse component of the applied field.

Certain noise issues can also be addressed by measurements of the  malefactor. For example, the longitudinal magnetic field can be  measured and used to correct transverse fluctuations if such fluctuations are caused by instabilities in the currents used to provide  $\vec{B}_{\text{ext}}$. Similarly, measurements of mechanical motions between the sample and the source of $\vec{B}_{\text{ext}}$ can be used to remove the effect of these motions on the magnetic field. Time varying external magnetic fields are also a source of noise and the sample must be screened from them. This can be done using superconducting shields which have been demonstrated to reduce ac field noise by over $10^{13}$ \cite{GPB}.

\subsection{Magnetization Noise}

The intrinsic magnetization noise of the sample is due to the random flips of the nuclear spins. A model for this spin noise was presented in \cite{SpinNoise}, and its power spectrum $S\left(\omega\right)$ was found to be
 \begin{equation}
\label{eqn: magnetization noise}
S\left(f \right) = \frac{1}{8} \left(\frac{T_2}{1 + T_2^2 \left(2 \pi f - 4 \pi \mu B\right)^2}\right).
\end{equation}
The spin noise is peaked around the Larmor frequency $2 \mu B$ and has a bandwidth $\frac{1}{T_2}$.
This spectrum was used to estimate the magnetization noise as in Fig 2. 
The  model of \cite{SpinNoise} is an approximation to the noise in a real material. But, the parametric features of this model such as the noise peak at the Larmor frequency with a bandwidth set by $T_2$ and its dependence on sample volume are expected to be realized in a real material \cite{Budker2}. This model of the intrinsic sample-magnetization noise is consistent with the expectations of the fluctuation-dissipation theorem and it has been experimentally confirmed in systems in equilibrium  \cite{ClarkeNoise, Staudacher}. Since the experimental frequency and bandwidth are always much greater than $1/T_1$, the magnetization noise at these frequencies is dominated by the spin-spin magnetic dipole $T_2$-relaxation processes. These local processes are unaffected by the polarization of the sample and hence this is a good estimate of the magnetization noise. 

We integrate this noise over the frequency bandwidth being measured (around any particular axion mass):
\begin{equation}
I = \int_{2 \mu_N B + \delta f - \frac{1}{4 \pi T_b}}^{2 \mu_N B + \delta f + \frac{1}{4 \pi T_b}} S \left( f \right) df,
\end{equation}
where $\delta f \sim 1 / \tau$ is the offset of the center of the axion signal from the Larmor frequency, and $T_b$ is defined to be the ``signal bandwidth time" so that $ \frac{1}{2 \pi T_b}$ is the bandwidth of the signal region being searched in.  $T_b$ is taken to be the smaller of the axion coherence time $\tau_a$ and $T_2$ for the measurement because whichever of these times is smaller defines the bandwidth in which we are looking for the signal.  Note that $T_2$ cannot be longer than the integration time at a particular frequency, though this requirement is not relevant given that the longest $T_2$ we have chosen to plot in Fig.~2 
 is $T_2 = 1$ s which is always shorter than the integration time at any frequency.

Then the magnetic field noise is
\begin{equation}
B_\text{noise} \approx \mu \, \sqrt{\frac{n}{V} \, I},
\end{equation}
where $V$ is the volume of the sample.  We convert this to the limit on $g_d$ exactly as done above for the magnetization signal, using Eqn.~\eqref{eqn: magnetization signal}.  In particular, we take the same scalings with time.  So the noise improves linearly in time up to the smaller of $T_2$ and $\tau_a$, then like the square root up to $\tau_a$ (if $\tau_a > T_2$), then like the fourth root up to the full integration time at this axion mass.

Note that there are cases where $T_2 > \tau_a$ (at the larger axion masses).  We are still allowed to take such a large $T_2$ even though it is longer than the time of an individual measurement $\tau_a$ so long as it is shorter than the full integration time at this particular axion mass.  This is because physically one may keep running the experiment and measuring the transverse magnetization of the sample without repolarizing it up to a time $T_1$.  This helps keep the magnetization noise curve lower at the higher axion masses.  This is not necessary for the experiment described here because even if we limited $T_2$ to be no longer than $\tau_a$ the magnetization noise curve would still be below the magnetometer noise curve in Fig.~2 
(and of course would not affect the magnetometer noise curves).  However, this strategy would be useful if for example magnetometers improved and the magnetometer noise was lowered.  The limit set by the magnetization noise would still allow observation of the QCD axion over the entire accessible region with frequencies $\lesssim 10^8$ Hz.

\subsection{Magnetometer Noise}

The transverse magnetization produced by the sample can be measured with a Superconducting Quantum Interference Device (SQUID) magnetometer. The magnetic flux from the sample is collected by a pickup coil (see Fig.~1) that is inductively coupled to a SQUID. The SQUID signal is proportional to the magnetic flux through the SQUID, which can be expressed in terms of sample magnetization $M$ as
\begin{align}
\label{eq:S1}
\Phi_{sq} = 4\pi M\left[ \zeta A\frac{NM_{in}}{L_{in}+L_p^{(N)}}\right],
\end{align}
where $\zeta$ is the sample demagnetization factor~\cite{Sushkov2009}, $A$ is the pickup loop area (matched to sample area), $N$ is the number of pickup loop turns (which is chosen to optimize the coupling to the SQUID), $M_{in}$ is the SQUID-input coil mutual inductance, $L_{in}$ is the SQUID input coil self-inductance, and $L_p^{(N)}$ is the $N$-turn pickup coil self-inductance, which is roughly proportional to the length of the pickup coil wire. We use the following parameters of a commercially-available SQUID: $M_{in}=10$~nH, $L_{in}=1.5$~$\mu$H. If the sample area $A=80$~cm$^2$ (which corresponds to a cylindrically-shaped sample of radius 5~cm), the demagnetization factor $\zeta \approx 0.7$, and the optimal number of pickup loop turns is $N=2$. These parameters set the value of the expression in the square brackets in Eq.~(\ref{eq:S1}), which we call the ``effective sample area'' $A_{eff}\approx 0.3$~cm$^2$.

The typical white noise level of a commercial SQUID is $\delta\Phi_{sq}\approx 1$~$\mu\Phi_0/\sqrt{\mathrm{Hz}}$, where $\Phi_0$ is the quantum of magnetic flux. Using the above parameters, we convert this to an effective magnetic field noise level of $\delta B_{sq}\approx 0.1$~fT$/\sqrt{\mathrm{Hz}}$, which is the value used in the text.


\subsection{Sensitivity Scaling with Averaging Time}

Consider a sinusiodally-varying magnetic field signal with frequency $f_0$ and amplitude $B_0$
\begin{align}
\label{eq:B1}
B(t) = B_0\sin\left[ 2\pi f_0 t + \phi(t)\right] + B_n(t),
\end{align}
where $B_n(t)$ is  noise, which we assume to be white, with power spectral density $S_n$. The signal has a phase coherence time $\tau$; we model this by assuming that the phase $\phi(t)$ remains constant at short times, and executes random jumps, separated by time $\tau$, to values uniformly distributed within the range 0 to $2\pi$. Ignoring factors of order unity, we derive signal-to-noise scaling with averaging time $T$, for two cases: 1) $T<\tau$, 2) $T>\tau$.

Consider the function
\begin{align}
\label{eq:B2}
P(f) = \frac{1}{\sqrt{T}}\int_0^T B(t) \sin(2\pi f t) dt = P_0(f) + P_n(f).
\end{align}
This is a Fourier transform, normalized in such a way so as to ensure that the noise energies $|P^2|$
add over consecutive time intervals; $P_0$ and $P_n$ correspond to the signal and the noise transforms.

In the limit $T\rightarrow\infty$, $|P_n(f)|^2\rightarrow S_n$~\cite{Champeney}. However, $P_n(f)$ is not a smooth function, each of the independent Fourier components, separated by frequency intervals $1/T$, is randomly distributed according to a Gaussian distribution centered at zero and with variance $S_n$.
From now on we shall focus on the quantity $|P(f)|^2 = |P_0(f)|^2 + |P_n(f)|^2$. Both the expectation value and the standard deviation of $|P_n(f)|^2$ are equal to $S_n$. Now let us compute $|P_0(f)|^2$.

1) The signal is phase coherent: $T<\tau$. Provided $T\gg 1/f_0$, the Fourier transform vanishes everywhere except at $f=f_0$, where (ignoring factors of order unity) $|P_0(f_0)|^2 = B_0^2 T$. The measurement sensitivity $B_s$ is the value of $B_0$ that satisfies $|P_0(f_0)|^2 = |P_n(f_0)|^2$, which yields
\begin{align}
\label{eq:B3}
B_s = \sqrt{S_n}T^{-1/2}.
\end{align}

2) The measurement time is longer than signal coherence time: $T>\tau$. The key point is that the signal Fourier transform $|P_0(f)|^2$ is no longer a ``delta function'', but now has a linewidth $1/\tau$. The peak amplitude $|P_0(f_0)|^2$ no longer grows with measurement time, but is a constant. We estimate this amplitude by breaking up the time-integral in Eq.~(\ref{eq:B2}) into $T/\tau$ pieces of duration $\tau$, and adding them in quadrature, since they have random phase pre-factors. Each piece contributes $B_0^2\tau^2$, thus the total value of the integral is $|P_0(f_0)|^2 = B_0^2\tau$.

If we extracted the measurement sensitivity as in case (1), by comparing $|P_0(f_0)|^2$ and $|P_n(f_0)|^2$, then we would find that the sensitivity does not improve with measurement time. However, $|P_0(f)|^2$ now has a linewidth $1/\tau$, which means we now have a signal not just at $f_0$, but at many frequencies around $f_0$, the number of independent frequency points is $T/\tau$. We now compare $|P_0(f)|^2$ and $|P_n(f)|^2$ for each of these frequency points, or fit a lineshape to $|P(f)|^2$, given noise in $|P_n(f)|^2$. As noted above, the scatter of the noise transform points $|P_n(f)|^2$ is $S_n$, and now we can effectively average $T/\tau$ such points near $f_0$. Thus the measurement sensitivity $B_s$ is the value of $B_0$ that satisfies $|P_0(f_0)|^2 = S_n/\sqrt{T/\tau}$, which yields
\begin{align}
\label{eq:B4}
B_s = \sqrt{S_n}(\tau T)^{-1/4}.
\end{align}

We have the usual scaling $B_s\propto T^{-1/2}$ as long as the signal is phase coherent, but beyond the coherence time the sensitivity scales as $T^{-1/4}$.

The magnetometer limit to the sensitivity is determined by the ratio between the signal size in
Eq.~\eqref{eqn: magnetization signal} and the magnetometer sensitivity.
The signal will increase linearly in time up to the minimum of $T_2$ and $\tau_a$.  The magnetometer noise goes as $\sim \frac{10^{-16}}{\sqrt{2 t}} \frac{\text{T}}{\sqrt{\Hz}}$, so the ratio improves as $\propto t^{\frac{3}{2}}$.  Then if $T_2$ is shorter than $\tau_a$, this ratio will increase $\propto \sqrt{t}$ up to $\tau_a$.  From then on it will increase as $t^\frac{1}{4}$, similar to, for example, \cite{Allen:1997ad}).
Using the above scalings with integration time $t$ we find the sensitivity curves in Fig.~\ref{Fig:EDM}.

\end{document}